\documentclass[10pt]{article} 
\def\sp{\vskip 3 true mm \noindent}
\def\p{\par\noindent}

\def \ot{\p \hbox to 8 cm{\hfill}}
\def \otb{\p \hbox to 2cm{\hfill}}

\def \r{\rightarrow}

\def\ol{\overline}
\def\ul{\underline}

\def\Xul{\underline{X}}

\def\eol{\overline{e}}

\def\sol{\overline{s}}

\def\Aol{\overline{A}}
\def\Bol{\overline{B}}

\def \5{[\hskip 5mm]}
\def\sqr#1#2{{\vcenter{\vbox{\hrule height.#2pt
        \hbox{\vrule width.#2pt height#1pt \kern#1pt
	   \vrule width.#2pt}
        \hrule height.#2pt}}}}
\def\dqed{\mathchoice\sqr65\sqr65\sqr{5.1}4\sqr{4.5}4}
\def\qed{$\dqed$}
\font\msbmten=msbm10 		%% Inladen van nieuwe fonts in TeX
\font\msbmseven=msbm7		%% In drie formaten, voor text-size
\font\msbmfive=msbm5		%% scriptsize en script-script-size

\newfam\amssym
\textfont\amssym=\msbmten	
\scriptfont\amssym=\msbmseven
\scriptscriptfont\amssym=\msbmfive

\begingroup
\multiply\count0 by "100\advance \count0 by '157
\global\mathchardef\rtimes=\count0
\endgroup

\newcounter{enumrom}
\newenvironment{enumrom}{\begin{list}{\roman{enumrom})}{\usecounter{enumrom}
\setlength{\rightmargin}{\leftmargin}}}{\end{list}}

\usepackage[all,v2]{xy}
\LaTeXdiagrams
\usepackage{amssymb}
\begin{document}
\title{Dequantization of Noncommutative Spaces\\
and Dynamical Noncommutative Geometry}
\author{
Freddy Van Oystaeyen\\
Department of Mathematics \& Computer Science\\
University of Antwerp\\
Middelheimcampus\\
Middelheimlaan 1, 2020 Antwerpen, Belgium\\
e-mail : fred.vanoystaeyen@ua.ac.be}
%\date
\maketitle

\begin{abstract}
	The purely mathematical root of the dequantization constructions is the quest for a sheafification needed for presheaves on a noncommutative space. The moment space is constructed as a commutative space, approximating the noncommutative space appearing as a dynamical space, via a stringwise construction. The main result phrased is purely mathematical, i.e. the noncommutative stalks of some sheaf on the noncommutative space can be identified to stalks of some sheaf associated to it on the commutative geometry (topology) of the moment space. This may be seen as a (partial) inverse to the deformation--quantization idea, but in fact with a much more precise behaviour of stalks of sheaves. The method, based on minimal axiomatics necessary to rephrase continuity  principles in terms of partial order (noncommutative topology) exclusively, leads to the appearance of objects like strings and (M--)branes. Also, spectral families and observables may be defined and studied as separated filtrations on the noncommutative topology! We highlight the relation with pseudo--places and generalized valuation theory. Finally, we hint at a new notion of ``space'' as a dynamical system of noncommutative topologies with the same commutative shadow (4 dimensional space--time if you wish) and variable (but isomorphic) moment spaces, which for special choices may be thought of as a higher dimensional brane--space.
\end{abstract}

\section*{Introduction}
Noncommutative spaces obtained by quantization-deformation may to some extent
be traced back to the base space they have been deformed from.  A
noncommutative topology in the sense of [VO1] has a so-called commutative
shadow that turns out to be a lattice.  However, the lack of geometric points
in a noncommutative space contrasts the set-theory based commutative geometry
with its function theory and corresponding analysis.  Looking for a converse
construction for quantization, a better commutative approximation of a
noncommutative space is needed; this construction should allow tracing of
physical aspects like observables, spectral families etc... 
\par
Different notions of noncommutative space have been introduced e.g.
noncommutative manifolds [Co], noncommutative ``quantized'' algebras [ATV],
[VdB], general quantization-deformations [Ko].  In these theories the actual
geometric object is left as a virtual object and results deal usually with
noncommutaive algebras or categories thought of as rings of functions or
categories of modules.
\par
In the author's approach, started in [VO1], a noncommutative topology is at
hand and then sheaf theory should to some extent replace function theory.
Applied to Physics this model seems to allow some explanation of quantum
phenomena but for explicite calculations and geometric reasoning a
commutative model is unavoidable because at some point coordinatization and
the use of real or complex numbers becomes necessary.  In the model we propose
these commutative techniques may be carried out in the dequantized space.
The construction of this commutative approximation of noncommutative space
derives from a dynamical version of noncommutative geometry that may also be
seen as a noncommutaive topological blow-up and blow-down construction.
Restricting assumptions to bare necessities one finds that all structural
properties only depend on a few intuitive axioms at the level of ordered
(partial) structures.  Respecting this philosophy we started from a totally
ordered set (not even necessarily a group), $T$ and suitably connected
noncommutative spaces $\Lambda_t, t\in T$.  A stringwise construction then
yields a new spectral space, called the {\bf moment space}, at each $t\in T$,
with its topology in the classical sense i.e. a commutative space.  
\par
Sheaves on lattices were already termed quantum sheaves, but now we view 
(pre-)sheaves on noncommutative topologies fitting together in a kind of 
dynamical geometry and connect them to sheaves on the moment spaces.  The idea
that commutative moment spaces are good approximations of noncommutative
spaces is reflected in the main result : the stalks at points of the moment
space at $t\in T$ equal stalks at a point at some $\Lambda_{t'}$, $t'\in T$
for some $t'$ in a prescribed closed $T$-interval containing $t$.  The
philosophy here is that there may not exist enough geometric points in the
noncommutative geometry $(\Lambda_t)$ at $t\in T$ but it holds in the moment 
space at $t$
because it encodes dynamical information in some $T$-interval containing
$t$.  The construction is abstract and rather general, but when looking for
applications in Physics one is only interested in one case ... reality ?  Then
one would think of $T$ as a multidimensional irreversible ``time'' with
``dimension'' big enough to make up for the missing geometruc points in the
noncommutative space $\Lambda_t$ at $t\in T$, i.e. the $T$-dimension explains
the difference in dimension between the noncommutative space, or the moment
space, and that of its space of points (the commutative shadow).  Whereas the
commutative shadow represents the abstract space where the mathematics takes
place, e.g. space-time, the moment space represents the space where the
mathematics of observed and measured items takes place, and because observing
or measuring takes time the latter space has to encode dynamical data.  Stated
as a slogan : ``Observation Creates Space from Time''.  Moreover, when
observing a smallest possible entity, we would like to think of this as an
observed point i.e. a point in the moment space, at $t$ say.  Tracing this via
the dynamical noncommutative geometry to the commutative shadow (thinking of
this as the fixed space we calculate in) the observed point appears as a
string (it may be open or closed).  On the other hand the point in the moment
space is via a string of temporal points (not geometric points) in the
noncommutative space also traceable to a possibly ``higher dimensional string''
in the moment space, in fact a tube-like string connecting opens in (the
topologies of the noncommutative spaces blown down to) the spectral topology
of the moment space (also identified to one commutative space for example).
Of course this suggests that a version of $M$-theory already appears in the
mathematical formalism explaining the transfer between commutative shadow and
moment space.
\par
Moreover the incertainty principle also marks the difference between the two
commutative spaces, in some sense  expressing a quantum-commutativity of the
noncommutative spaces, for example phrased in sheaf theoretical terms.  
\par
We have aimed at a rather self contained presentation of basic facts and
definitions of the new mathematical objects in particular ... generalized
Stone spaces and noncommutative topology.
\par
The methods of this paper can be adapted to other types of noncommutative
spaces e.g.quantales,...  . Further generalization, e.g. to noncommutative
Grothendieck topologies, has not been included, this would be an exercise
using some material from [VO1].  We include some comments about specific
examples and further mathematical elaborations using deformations of affine
spaces or other nice classical (algebraic) varieties is left as work in
progress.
\section{Preliminaries on Noncommutative Topology}
We consider (partially ordered) $\Lambda$ with $0$ and $1$.  When $\Lambda$ is
equipped with operations $\wedge$ and $\vee$, we say that $\Lambda$ is a
noncommutative topology if the following axioms hold~:
\begin{enumrom}
\item
for $x,y\in\Lambda, x\wedge y\le x$ and $x\wedge y\le y$ and $x\wedge
1=1\wedge x=x$, $x\wedge 0=0\wedge x=0$, moreover $x\wedge\ldots \wedge x=0$
if and only if $x=0$.
\item
For $x,y,z\in\Lambda$, $x\wedge y\wedge z=(x\wedge y)\wedge z=x\wedge (y\wedge
z)$, and if $x\le y$ then $z\wedge x\le z\wedge y, x\wedge z\le y\wedge z$.
\item
Properties similar to i. and ii. with respect to $\vee$, in particular $x\vee
x\vee\ldots\vee x=1$ if and only if $x=1$.
\item
Let ${\rm
id}_{\wedge}(\Lambda)=\{\lambda\in\Lambda,\lambda\wedge\lambda=\lambda\}$; for
$x\in{\rm id}_{\wedge}(\Lambda)$ and $x\le z$ we have~: $x\vee(x\wedge z)\le
(x\vee x)\wedge z$, $x\vee (z\wedge x)\le (x\vee z)\wedge x$.
\item
For $x\in\Lambda$ and $\lambda_1,\ldots,\lambda_n\in\Lambda$ such that
$1=\lambda_1\vee\ldots\vee\lambda_n$, we have
$x=(x\wedge\lambda_1)\vee\ldots\vee(x\wedge\lambda_n)=(\lambda_1\wedge
x)\vee\ldots\vee(\lambda_n\wedge x)$.
\end{enumrom}
\par
There are left (right) versions of this definition as introduced in [VO.1].,
but we do not go into this here.  In fact we restrict attention to the
situation where $\vee$ is a commutative operation and $\Lambda$ is
$\vee$-complete i.e. for an arbitrary family ${\cal F}$ of elements in
$\Lambda$, $\vee{\cal F}$ exists in $\Lambda$, where $\vee{\cal F}$ is
characterized by the property~: $\lambda\le \vee{\cal F}$ for all
$\lambda\in{\cal F}$ and if $\lambda\le\mu$ for all $\lambda\in{\cal F}$ then
$\vee{\cal F}\le\mu$.\par
In case only v) is dropped we call $\Lambda$ a {\bf skew topology}; this is
sometimes interesting e.g. the lattice $L(H)$ of a Hilbert space $H$ is a skew
topology and condition v) does not hold !  Here $L(H)$ is the lattice of
closed linear subspaces of $H$ with respect to intersection and closure of the
sum.  
The restriction to abelian $\vee$ (most interesting examples are like this)
entails that ${\rm id}_{\wedge}(\Lambda)$ is closed under the operation $\vee$.
Moreover on ${\rm id}_{\wedge}(\Lambda)$ we introduce a new operation
\d{$\wedge$} defined by $\sigma$\d{$\wedge$}$\tau=\vee\{\gamma\in{\rm
id}_{\wedge}(\Lambda),\gamma\le\sigma\wedge\tau\}$ for $\sigma,\tau\in{\rm
id}_{\wedge}(\Lambda)$.  Let us write $SL(\Lambda)$ for the set ${\rm
id}_{\wedge}(\Lambda)$ with the operations \d{$\wedge$} and $\vee$; then
$SL(\Lambda)$ is easily checked to be again a skew topology (with $\vee$
commutative and being $\vee$-complete).
\subsection{Lemma (cf. [VO.1] or [VO.2] 2.2.3. and 2.2.5)}
If $\Lambda$ is a skew topology $\vee$-complete with respect to a commutative
operation $\vee$, then $SL(\Lambda)$ is a lattice satisfying the modular
inequality.
\par
A subset $X \subset \Lambda$ is {\bf directed} if for every $x,y\in X$ there
is a $z\in X$ such that $z\le x, z\le y$; we say that $X$ is a {\bf filter} in
$\Lambda$ if it is directed and for $x\in X$, $x\le y$ yields $y\in X$.  Two
directed sets $A$ and $B$ in $\Lambda$ are equivalent if they define the same
filter $\Aol=\Bol$ where for any directed set $A$ we put 
$\Aol=\{\lambda\in\Lambda$, there is an $a\in A$ such that $a\le\lambda\}$.
Let $D(\Lambda)$ be the set of directed subsets in $\Lambda$ and we write
$A\sim B$ of the directed subsets $A$ and $B$ are equivalent, we write $[A]$
for the equivalence class of $A$ and let $C(\Lambda)$ be the set of classes of
directed subsets of $\Lambda$.  We introduce a partial order in $C(\Lambda)$
by putting $A\le B$ if $\Bol\subset \Aol$ and $[A]\le [B]$ induced by the
partial order on ${\cal D}(\Lambda)$ defined by the foregoing.  For $A$ and
$B$ in ${\cal D}(\Lambda)$ define $A\dot{\wedge} B=\{a\wedge b, a\in A, b\in
B\}, A\dot{\vee}B=\{a\vee b, a\in A, b\in B\}$ and~: $[A]\wedge
[B]=[A\dot{\wedge}B],[A]\vee[B]=[A\dot{\vee}B]$.
\subsection{Lemma}
If $\Lambda$ is a skew topology, resp. a noncommutative topology, then so is
$C(\Lambda)$ with respect to the partial order and operations $\wedge$ and
$\vee$ as defined above.  The canonical map $\Lambda\r
C(\Lambda),\lambda\mapsto [\{\lambda\}]$ is a map of skew topologies.
\par
We simplify notations by writing $[\{\lambda\}]=[\lambda]$ and call such an
element {\bf classical} in $C(\Lambda)$.
\par
A directed set $A$ in $\Lambda$ is {\bf idempotently directed} if for every
$a\in A$ there exists an $a'\in A\cap {\rm id}_{\wedge}(\Lambda)$ such that
$a'\le a$; in this case $[A]\in i_{\wedge}(C(\Lambda))$ but these elements of
$C(\Lambda)$ may be thought of as obtained from a directed set having a
cofinal subset of ``commutative'' opens (the idempotents belonging to the
commutative shadow $S((\Lambda))$.  We write ${\rm Id}_{\wedge}(C(\Lambda))$
for the subset of ${\rm id}_{\wedge}(C(\Lambda))$ consisting of the classes of
idempotently directed subsets of $\Lambda$; the elements $[A]$ of ${\rm
Id}_{\wedge}({\rm Id}_{\wedge}(C(\Lambda))$ are called {\bf strongly
idempotents}.  We identify $\Lambda$ and the image of $\Lambda\r C(\Lambda)$,
then observe that ${\rm Id}_{\wedge}(C(\Lambda))\cap \Lambda={\rm
id}_{\wedge}(\Lambda)$.  We shall write $\prod(\Lambda)$ for the skew
(noncommutative) topology obtained by talking $\wedge$-finite bracketed
expression $P(\wedge,\vee,x_i)$ in terms of strong idempotents $x_i\in{\rm
Id}_{\wedge}(C(\Lambda))$; similarly we write $T(\Lambda)$ for the skew
(noncommutative) topology obtained by taking $\wedge$-finite bracketed
expressions $p(\wedge,\vee,\lambda_i)$ in idempotents $\lambda_i\in {\rm
id}_{\wedge}(\Lambda)$.
\par
It is not hard to verify~: $\prod(\Lambda)=\prod(T(\Lambda))$.  Moreover
$C(T(\Lambda))$ satifies the same axions (i. to v.) as $\Lambda$ and
$T(\Lambda)$ but with respect to ${\rm Id}_{\wedge}(C(\Lambda))$.  The
``strong'' commutative shadow of $\prod$ is obtained by restricting
\d{$\wedge$} on ${\rm id}_{\wedge}(C(\Lambda))$ to ${\rm
Id}_{\wedge}(C(\Lambda))$ and viewing $SL_{s}(\prod)$ as the lattice structure
induced on ${\rm Id}_{\wedge}(C(\Lambda))$.
\subsection{Lemma}
If $\Lambda$ is a $\vee$-complete noncommutative topology such that $\vee$ is
commutative then~: $SL_s(\prod)=C(SL(\Lambda))$.
\subsection{Definition.  Generalized Stone Topology}
Consider a skew topology $\Lambda$ and $C(\Lambda)$.  For $\lambda\in\Lambda$,
let $O_{\lambda}\subset C(\Lambda)$ be given by
$O_{\lambda}=\{[A],\lambda\in\Aol\}$.  It is very easy to verify~:
$O_{\lambda\wedge\mu}\subset O_{\lambda}\cap O_{\mu},O_{\lambda\vee\mu}\supset
O_{\lambda}\cup O_{\mu}$, hence the $O_{\lambda}$ define a basis for a
topology on $C(\Lambda)$ termed~: generalized Stone topology.  This definition
obtains a more classical meaning when related to suitable point-spectra
constructed in $C(\Lambda)$.
\par
We say that $[A]$ in $C(\Lambda)$ is a {\bf minimal point of $\Lambda$} if
$\Aol$ is a maximal filter, i.e. $\Aol\neq \Lambda$ but if $\Aol\subsetneqq B$
where $B$ is a filter then $B=\Lambda$; it follows that a minimal point is
necessarily in ${\rm id}_{\wedge}(C(\Lambda))$ and it is indeed a minimal
nonzero element of the poset $C(\Lambda)$.  An {\bf irreducible point} $[A]$
of $\Lambda$ is characterized by either one of the followingequivalent
properties~:
\begin{enumrom}
\item
$[A]\le \vee\{[A_{\alpha}],\alpha\in{\cal A}\}$ yields $[A]\le[A_{\alpha}]$
for some $\alpha\in{\cal A}$.
\item
If $\vee\{\lambda_{\alpha},\alpha\in{\cal A}\}\in\Aol$ then
$\lambda_{\alpha}\in \Aol$ for some $\alpha\in{\cal A}$.
\end{enumrom}
More general types of points may be considered, e.g.the elements of a
so-called quantum-basis, cf [VO.2], but we need not go into this here.  Under
some suitable condition often (but not always) present in examples the
irreducible points in ${\rm Id}_{\wedge}(C(\Lambda))$ are exactly those that
are $\vee$-irreducible in $C(\Lambda)$ (e.g. if $\Lambda$ satisfies the weak
FDI property, cf. [VO.1], Proposition 5.9.).
\par
Define the (irreducible) {\bf point-spectrum} by putting~:  ${\rm
Sp}(\Lambda)=\{[p],[p]$ an (irreducible) point of $\Lambda\}$.  Put
$p(\lambda)=\{[p]\in {\rm Sp}(\Lambda),[p]\le [\lambda]\}$ for
$\lambda\in\Lambda$, then $p(\lambda\wedge\mu)\subset p(\lambda)\cap p(\mu),
p(\lambda\vee \mu)=p(\lambda)\cup p(\mu)$.  Thus the $p(\lambda)$ define a
basis for a topology on ${\rm Sp}(\lambda)$ called the point-topology.  Write
$SP(\Lambda)$ for $Sp(\Lambda)\cap {\rm Id}_{\wedge}(C(\Lambda))$ and refer to 
this as the {\bf Point-spectrum} (capital $P$).  For $\lambda\in \Lambda$ we 
consider
$P(\lambda)=\{[P]\in {\rm SP}(\Lambda),[P]\le [\lambda]\}$ and this indices
the Point-topology on ${\rm SP}(\Lambda)$; this time we even have
$P(\lambda\wedge\mu)=P(\lambda)\cap P(\mu)$ and this time we even have
$P(\lambda\wedge\mu)=P(\lambda)\cap P(\mu)$ and this defines a topology on
${\rm SP}({\rm SL}(C(\Lambda)))$.  Similar constructions may be applied to the
minimal point-spectrum $Q{\rm Sp}(\Lambda)$ on ${\rm Sp}(\Lambda)$ the
generalized Stone topology is nothing but the point-topology.  In the
foregoing $\Lambda$ may be replaced by $T(\Lambda)$ or $\prod(\Lambda)$ with
topologies induced by the generalized Stone topology on the point spectra
always again being called~: generalized Stone topology.  Finally the
generalized Stone topology can also be defined on the commutative shadow ${\rm
SL}(\Lambda)$, which is a modular lattice, and then the topology  induced on
$Q{\rm SP}({\rm SL}(\Lambda))$ is exactly the Stone topology of the Stone
space of ${\rm SL}(\Lambda)$.
\par
In the special case $\Lambda=L(H)$ (only a skew topology) the generalized
Stone space defined on $Q{\rm SP}(L(H))$ is exactly the classical Stone space
as used in Gelfand duality theory for $L(H)$.  Warning~: $L(H)$ does not
satisfy the axiom v. and whereas $Q{\rm SP}(L(H))$ is rather big, ${\rm
Sp}(L(H))={\rm SP}(L(H))$ is empty !  Moreover over $L(H)$ there are no
sheaves but there will be many sheaves over $C(L(H))$ making sheaffication of
a separated presheaf over $L(H)$ possible over $Q{\rm SP}(L(H))$.
\par
The foregoing fixes a context for results in the sequel, however the methods
in Sections 2-5 are rather generic and can be applied to other notions of
noncommutative spaces.
\section{Dynamical Noncommutative Topology}
The realization of a relation between the static and the dynamic goes far
back, at least to D'Alembert but the notion that the first may be used to
study the latter may be not completely well founded.  Almost existentialistic
problems defy the correctness of a mathematical description of so-called
physical reality.  So we choose for a converse approach constructing space as
a dynamical noncommutative topological space and defining geometrical objects
as existing over some parameter-intervals.  Noncommutative continuity is
introduced via the variation of an external parameter in a totally ordered set
$T$ (if one wants to consider this as a kind of time, fine... but then this
time is an index, not a dimension).  This is philosophically satisfying
because momentary observations which are only abstractly possible (real
measurement takes time) put us in the discrete-versus-continuous situation of
noncommutative geometry !
\par
Let $T$ be a totally ordered set and for every $t\in T$ we give a
noncommutative space $\Lambda_t$.  This can have several meanings, in the
sequel we take this to mean that $\Lambda_t$ is the generalized Stone space
constructed on $C(X_t)$ for some skew topology $X_t$ as in Section 1.  This is
just to fix ideas, in fact one could just as well restrict to topologies
induced by the genaralized Stone topology on point spectra of any type, see
also Section 1 or [VO1], or take pattern topologies as introduced in [VO1,2],
or go to other theories and take quantales etc...  For $t\le t'$ in $T$ we
have $\varphi_{tt'}:\Lambda_t\r \Lambda_{t'}$ which are poset maps respecting
$\wedge$ and $\vee$; when $t=t'$ we take $\varphi_{tt}=1_{\Lambda_t}$ to be
the identity of $\Lambda_t$ and when $t\le t'\le t''$ then we demand that
$\varphi_{t't''}\circ \varphi_{tt'}=\varphi_{tt''}$, where notation for
composition of maps is conventional i.e. writing the one acting last frst.  If
$A_t\subset \Lambda_t$ is a directed set then $\varphi_{tt'}(A_t)\subset
\Lambda_{t'}$, for $t\le t'$, is again a directed set.
\par It is easily verified that if we start from a system
$\{X_t,\Psi_{tt'},T\}$ defined as above, we obtain a similar system
$\{C(X_t),\Psi^e_{tt'},T\}$ where $\Psi^e_{tt'}:C(X_t)\r C(X_{t'})$ is given
by putting~: $\Psi^e_{tt'}([A])=[\Psi_{tt'}(A)]$, for $[A]\in C(X_t)$ and $t\in
t'$ in $T$.  In case it is interesting to view $\Lambda_t$ as coming from some
$X_t$ via $C(X_t)$ we may restrict attention to systems given by 
$\varphi_{tt'}$,
$t\le t'$ in $T$, stemming from $\psi_{tt'}$ on $X_t$ by extension as
indicated above.  Note that not every system $\{C(X_t),\varphi_{tt'},T\}$ has
to derive from a system $\{X_t,\psi_{tt'},T\}$ in general.
\subsection{Lemma}
Any system of poset maps $\varphi_{tt'},t\le t'$ in $T$, defines a system of
poset maps $\varphi^e_{tt'},t\le t'$ in $T$.  If the maps $\varphi_{tt'}$
respect the operations $\wedge$ and $\vee$ in the $\Lambda_t$ then so does
$\varphi^e_{tt'}$ for $C(\Lambda_t), t\in T$.  In this situation
$\varphi_{tt'}$ maps $\wedge$-idempotent elements of $\Lambda_t$ to
$\wedge$-idempotent elements of $\Lambda_{t'}$ (also $\vee$-idempotent to
$\vee$-idempotent) moreover if $[A_t]$ is strongly idempotent in
$C(\Lambda_t)$ then $[\varphi_{tt'}(A_t)]$ is a strongly idempotent element of
$C(\Lambda_{t'})$, for every $t\le t'$ in $T$,
\paragraph{Proof}
First statements follow obviously from~: for directed sets $A$ and $B$,
$$\displaylines{
\varphi^e_{tt'}([A]\wedge[B])=\varphi^e_{tt'}([A\dot{\wedge}B])
=[\varphi_{tt'}(A\dot{\wedge} B)]\hfill \cr
\hfill =[\varphi_{tt'}(A)\dot{\wedge}\varphi
_{tt'}(B)]=[\varphi_{tt'}(A)]\wedge [\varphi_{tt'}(B)]}$$
for $t\le t'$ in $T$.
Similar with respect to $\vee$, using $\dot{\vee}$.  In case
$\lambda\in\Lambda_t$ is idempotent in $\Lambda_t$ then
$\varphi_{tt'}(\lambda)\wedge
\varphi_{tt'}(\lambda)=\varphi_{tt'}(\lambda\wedge
\lambda)=\varphi_{tt'}(\lambda$ for $t\le t'$ in $T$.  Finally if $A$ is
idempotently directed look at $\varphi_{tt'}(a)$ for $a\in A_t$; by assumption
there exists some $\mu\in{\rm id}_{\wedge}(\Lambda_t)$ such that $\mu\le a$,
thus $\varphi_{tt'}(\mu)\le \varphi_{tt'}(a)$ and $\varphi_{tt'}(\mu)\in
{\rm id}_{\wedge}(\varphi_{tt'}(A_t))$, for $t\le t'$ in $T$.  Consequently
$\varphi_{tt'}(A_t)$ is idempotently directed too.  
\par
The skew topology $\prod_t$, introduced after Lemma 1.2. is called the pattern
topology of $X_t$, i.e. it is obtained by taking all $\wedge$-finite bracketed
expressions with respect to $\vee$ and $\wedge$ in the letters of
${\rm Id}_{\wedge}(C(\Lambda_t))$.
\subsection{Corollary}
The system $\{\Lambda_t,\varphi_{tt'}T\}$ induces a system
$\{\prod_t,\varphi_{tt'}|\prod_t,T\}$, satisfying the same properties, on the
pattern topologies.
\par
In general the $\varphi_{tt'},t\le t'$, do not map poits of $\Lambda_t$ to
points of $\Lambda_{t'}, t\le t'$, neither does $\varphi_{tt'}$ respect the
operation \d{$\wedge$} of the commutative shadow $SL(\Lambda_t)$, i.e. the
$\varphi_{tt'}$ do not necessarily induce a system on the commutatve shadows !
\subsection{Axioms for Dynamical Noncommutative Topology (DNT)}
A system $\{\Lambda_t,\varphi_{tt'},T\}$ is called a DNT if the following five
conditions are satisfied~:
\begin{itemize}
\item[DNT.1]
Writing $0$, resp. 1, for the minimal, resp. maximal element of $\Lambda_t$
(we shall assume these exist throughout) then $\varphi_{tt'}(0)=0$,
$\varphi_{tt'}(1)=1$ for all $t\le t'$ in $T$.
\item[DNT.2]
For all $t\in T, \varphi_{tt}=1_{\Lambda_t}$ and for $t\le t'\le t''$ in $T$,
$\varphi_{t't''}\circ \varphi_{tt'}=\varphi_{tt''}$.  Moreover, all
$\varphi_{tt'}$ preserve $\wedge$ and $\vee$.  Hence DNT.1. and 2. just
restate that $\{\Lambda_t,\varphi_{tt'},T\}$ is as before.
\item[DNT.3]
If for some $t\in T$, $0<x<y$ in $\Lambda_t$, then there is a $t<t_1$ in $T$
such that for $z_1\in \Lambda_{t_1}$ satisfying $\varphi_{tt_1}(x)<z_1<\varphi_{tt_1}(y)$ there is a $z\in \Lambda_t$, $x<z<y$, such that
$\varphi_{tt_1}(z)=z_1$.  
\item[DNT.4]
For every $t\in T$ and $0<x<z<y$ in $\Lambda_t$ there exist $t_1,t_2\in T$
such that $t_1<t<t_2$ and for every $t'\in ]t_1,t_2[$ we have either $t\le t'$
and $\varphi_{tt'}(x)<\varphi_{tt'}(z)<\varphi_{tt'}(y)$, or $t'\le t$ and
then if $x'< y'$ in $\Lambda_{t'}$ exist such that
$\varphi_{t't}(x')=x,\varphi_{t't}(y')=y$ then there also exist $z'$ in
$\Lambda_{t'}$ such that $x'<z'<y'$ and $\varphi_{t't}(z')=z$.  Taking the
special case $y=1$ and $y'=1$ then we see that a nontrivial strict relation in
$\Lambda_t$ is alive in an open $T$-interval containing $t$.
\item[DNT.5]
{\bf $T$-local unambiguity}.  In the situation of DNT.3, resp. DNT.4, the
$t_1\in T$, resp. $t_1$ and $t_2$, may be chosen such that $z\in \Lambda_t$ is
the unique element such that $\varphi_{tt_1}(z)=z_1$, resp. $x',y',z'$ in
$\Lambda_t$, are the unique elements such that $\varphi_{t't}(z')=z$,
$\varphi_{t't}(y')=y$, $\varphi_{t't}(x')=x$, or when $t\le t'$ the $x,y,z,$
are unique elements mapping to $\varphi_{tt'}(z),\varphi_{tt'}(y),y_{tt'}(x)$
resp.
\end{itemize}
Since we are able to take finite intersection of open $T$-intervals in the
totally ordered set $T$, we may extend the foregoing to finite chains of
$0<x_1<x_2<\ldots <x_n,n\ge 3$.
\par
Observe that the axioms allow that non-interacting elements, i.e. $x$ such
that $0 < x < 1$ is the only orderrelation it is involved in, may appear and
disappear momentarily.  here disappearing means going to 0 under all
$\varphi_{tt'}$, $t\le t'$, if $x\in\Lambda_t$.
\par
Very often properties studied are only preserved in some $T$-interval, in
particular this happens often when trying to derive a property of a related
system from another one that may be globally defined for $T$.  This leads to
an interesting phenomenon, already encoding some aspect of the moment-spaces
to be defined later.
\subsection{Definition of Observed Truth}
A statement in terms of finitely many ingredients of a DNT and depending on
parametrization by $t\in T$ is said to be an {\bf observed truth} at $t_0\in T$ 
if there is an open $T$-interval $]t_1,t_2[$ containing $t_0$, such that the
statement holds for parameter values in this interval.
\par It seems that mathematical statements about a DNT turn into ``observed
truth'' when one tries to check them in the commutative shadow, meaning on
the negative side that many global (over $T$) properties of a DNT cannot be
established globally over $T$ in the commutative world !
\par
The noncommutative topologies $\Lambda_t$ considered in the sequel will be
such that $\vee$ is commutative and $\vee$ of arbitrary families exist; in
fact one may restrict to so-called ``virtual topologies'' as introduced in
[VO2]; here we do not need to assume axiom ($\nu$) with respect to global
covers, we may want to restrict to this case when needed.  We refer to the
special case mentioned as a DVT.
\subsection{Proposition}
Let $\{\Lambda_t,\varphi_{tt'},T\}$ be a DVT and let $SL(\Lambda_t)$ be the
commutative shadow of $\Lambda_t$ with maps $\varphi_{tt'}:SL(\Lambda_t)\r
SL(\Lambda_{t'})$, $t\le t'$ in $T$, just being the restrictions of the
$\varphi_{tt'}$ (using same notation).  Then the statement that
$\{SL(\Lambda_t),\varphi_{tt'},T\}$ is a DVT too is an observed truth at every
$t_0\in T$.
\paragraph{Proof}
All $\varphi_{tt'}$ map $\wedge$-idempotents to $\wedge$-idempotents, cf.
Lemma 2.1., so DNT.1. is obvious.  For DNT.2 we have to check that
$\varphi_{tt'}$ preserves \d{$\wedge$} on ${\rm id}_{\wedge}(\Lambda_t)$,
since \d{$\vee$}$=\vee$ now we have nothing to check for \d{$\vee$}.  Look at
$t_0\in T$, $\varphi_{t_0t}:\Lambda_{t_0}\r \Lambda_t$ and $\sigma,\tau$ in
${\rm id}_{\wedge}(\Lambda_{t_0})$.  If $\sigma<\tau$ then
$\varphi_{t_0t}(\sigma)\le \varphi_{t_0t}(\tau)$ and
$\varphi_{t_0t}(\sigma)$\d{$\wedge$}$\varphi_{t_0t}(\tau)=\varphi_{t_0t}(\sigma)=\varphi_{t_0t}(\sigma$\d{$\wedge$}$\tau)$, interchange the role of $\sigma$
and $\tau$ in case $\tau <\sigma$.  So we may assume $\sigma$ and $\tau$ to be
incomparable.  Restricting $t$ to a suitable $T$-interval (DNT 5) we may
assume that $\varphi_{t_0t}(\sigma)\neq\varphi_{t_0t}(\tau)$.  Assume
$\varphi_{t_0t}(\sigma$\d{$\wedge$}$\tau)<\varphi_{t_0t}(\sigma)$\d{$\wedge$}
$\varphi_{t_0t}(\tau)$.
\par
If $\varphi_{t_0t}(\sigma$\d{$\wedge$}$\varphi_{t_0t}(\tau)=
\varphi_{t_0t}(\sigma_)$ (a similar argument will hold if $\sigma$ and $\tau$
are interchanged) then $\varphi_{t_0t}(\sigma)\le\varphi_{t_0t}(\tau)$, hence
$\varphi_{t_0t}(\sigma)<\varphi_{t_0t}(\tau)$.  Using DNT.5 again, taking $t$
close enough to $t_0$, we obtain $\sigma$\d{$\wedge$}$\tau <\sigma_1<\tau$
such that $\varphi_{t_0t}(\sigma$\d{$\wedge$}$\tau)<\varphi_{t_0t}(\sigma_1)
=\varphi_{t_0t}(\sigma)<\varphi_{t_0t}(\tau)$.  Passing to $[t_0,t]$ small
enough in order to have unambiguity for $\varphi_{t_0t}(\sigma)$, we arrive at
$\sigma_1=\sigma$, contradicting the incomparability of $\sigma$ and $\tau$.
Therefore we arrive at strict relations~:
$\varphi_{t_0t}(\sigma$\d{$\wedge$}$\tau)<\varphi_{t_0t}(\sigma)$\d{$\wedge$}
$\varphi_{t_0t}(\tau)<\varphi_{t_0t}(\sigma),\varphi_{t_0t}(\tau)$.  We may 
moreover assume (DNT.3) that $t$ is close enough to $t_0$ so that there is a
$z\in\Lambda_{t_0}$ such that $\sigma$\d{$\wedge$}$\tau<\sigma,\tau$ and
$\varphi_{t_0t}(z)=\varphi_{t_0t}(\sigma)$\d{$\wedge$}$\varphi_{t_0t}(\tau)$.
If $z$ is not $\wedge$-idempotent, then $\sigma$\d{$\wedge$}$\tau<z\wedge
z<\sigma$ would lead to $\varphi_{t_0t}(z\wedge
z)=\varphi_{t_0t}(\sigma)$\d{$\wedge$}$\sigma_{t_0t}(\tau)$ because
$\varphi_{t_0t}$ respects $\wedge$ and the latter is idempotent in
$\Lambda_t$; then $\varphi_{t_0t}(z)=\varphi_{t_0t}(z\wedge z)$ but the
unambiguity guaranteed by the choice of $t$ close enough to $t_0$ (DNT.5) then
yields $z=z\wedge z$ or $z\in {\rm id}_{\wedge}(\Lambda_{t_0})$.  Thus
$z=\sigma$\d{$\wedge$}$\tau$ by definition, a contradiction.  Consequently,
for $t$ in some small enough $T$-interval containing $t_0$ we have
obtained~:$\varphi_{t_0t}(\sigma$\d{$\wedge$}$\tau)=\varphi_{t_0t}(\sigma)$\d{$\wedge$}$\varphi_{t_0t}(\tau)$, thus DNT.2 is an observed truth.  To check
DNT.3, take $\sigma < \tau$ in ${\rm id}_{\wedge}(\Lambda_{t_1})$, $t<t_1$ such
that $z_1\in {\rm id}_{\wedge}(\Lambda_{t_1})$ exists such that we have
$\varphi_{tt_1}(\sigma)<z_1<\varphi_{tt_1}(\tau)$.  Now by DNT.3. for
$\{\Lambda_t,\varphi_{tt'},T\}$ there is a $z\in\Lambda_t, z < \tau$, such that
$\varphi_{tt_1}(z)=z_1$, and DNT.5 for $(\Lambda_t,\varphi_{tt'},T\}$, used as
in foregoing part of the proof, yields $\varphi_{tt_1}(z)=z_1$ with $z$ also
$\wedge$-idempotent in $\Lambda_{t}$, for $t_1$ close enough to $t$.  The proof
of DNT.4 follows in the same way and DNT.5 is equally obvious because
unambiguity in a suitable $T$-interval allows to pull back idempotency.
Therefore all DNT-axioms hold for $\{SL(\Lambda_t),\varphi_{tt'},\tau)$ in a
suitable $T$-interval, hence we have the observed truth statement that
$\{SL(\Lambda_t),\varphi_{tt'},T\}$ is DNT.\hfill\qed
\sp
Now fix a notion of point i.e. either minimal point or irreducible point as in
Section 1. We say that $\lambda_t\in \Lambda_t$ is a {\bf temporal point} if
$t\in ]t_0,t_1[$ such that for some $t'\in ]t_0,t_1[$ there is a point
$p_{t'}\in\Lambda_{t'}$ such that~: either $t\le t'$ and
$\varphi_{tt'}(\lambda_t)=p_{t'}$, or $t'\le t$ and
$\varphi_{t't}(p_{t'})=\lambda_t$; in the first case we say $\lambda_t$ is a
{\bf future point}, in the second case a {\bf past point}.  The system
$\{\Lambda_t,\varphi_{tt'},T\}$ is said to be {\bf temporally pointed} if for
every $t\in T$ and $\lambda_t,\in\Lambda_t$ there exists a family of
temporal points $\{p_{\alpha,t};\alpha\in{\cal T}\}$ in $\Lambda_t$ such
that $\lambda_t$ is covered by it, i.e.
$\lambda_t=\vee\{p_{\alpha,t},\alpha\in{\cal A}\}$.  Write $T{\cal
P}(\Lambda_t)$ for the set of temporal points of $\Lambda_t$, if we
write ${\rm Spec}(\Lambda_t)=\{p_{t'}$ point in $\Lambda_{t'},p_{t'}$ defines
a temporal point of $\Lambda_t\}$ then $T{\cal P}(\Lambda_t)$ may also be
written as $T{\rm Spec}(\Lambda_t)$ (note $T{\rm Spec}(\Lambda_t)$ is in
$\Lambda_t$ but ${\rm Spec}(\Lambda_t)$ not).
\par
We need to build in more ``continuity'' aspects in the DVT-axioms without
using functions or extra assumptions on $T$ e.g. that it should be a group.  A
temporary poited system $\{\Lambda_t,\varphi_{tt'},T\}$ is a {\bf space
continuum} if the following conditions hold~:
\begin{itemize}
\item[SC.1]
There is a minimal closed interval $T_t$ containing $t$ in $T$ such that
$T{\rm Spec}(\Lambda_t)$ has support in $I_t$.  The set of points in
$\Lambda_{t'}$ with $t'\in I_t$, representing temporary points in $\Lambda_t$
is then called the {\bf minimal spectrum} for $T{\rm Spec}(\Lambda_t)$,
denoted by ${\rm Spec}(\Lambda_t,I_t)$.
\item[SC.2]
For any open $T$-interval $I$ such that $I_t\subset I$ there exists an open
$T$-interval $I^*_t$ with $t\in I^*_t$, such that for $t'\in I^*_t$ we have
$I_{t'}\subset I$.
\item[SC.3]
For intervals $[t_1,t_2]$ and $[t_3,t_4]$ we write $[t_1,t_2]<[t_3,t_4]$ if
$t_1\le t_3$ and $t_2\le t_4$ (similarly for open intervals).  If $t\le t'$ in
$T$ then $I_t < I_{t'}$.  This provides an ``orientation'' of the variation of
the minimal spectra !
\item[SC.4] {\bf Local Preservation of Directed Sets} For given $t\le t'$ in
$I_t$ and any directed set $A_t$ in $\Lambda_t$, the subset $\{\gamma_t\in
A_t$, there exists $\xi_t<\gamma_t$ in $A_t$ such that
$\varphi_{tt'}(\xi_t)<\varphi_{tt'}(\gamma_t)\}$ is cofinal in $A_t$ (defines
the same limit $[A_t]$).  For $t''\le t$ in $I_t$ there is a directed set
$A_{t''}$ in $\Lambda_{t''}$ mapped by $\varphi_{t''t}$ to a cofinal subset of
$A_t$.
\end{itemize}
A subset $J$ of $T$ is {\bf relative open around $t\in T$} if it is
intersection of $I_t$ and an open $T$-interval.  For
$x=(\ldots,x_t,\ldots)\in\prod_{t\in T}\Lambda_t$ we put ${\rm sup}(x)=\{t\in
T,x_t\neq 0\}$.  We say that such an $x$ is {\bf topologically accessible} if
all $x_{t'}t\in {\rm sup}(x)$, are classical i.e. $x_t=[\chi_t]$ (for some
$\chi_t\in X_t$ and $\Lambda_t=C(X_t)$.  In case we do not consider
$\Lambda_t$ as coming from some $X_t$ the condition becomes void.
An $x$ as before is said to be {\bf $t$-accessible} if ${\rm sup}(x)=J$ is
relative open around $t$ and for all $t' \le t''$ in $J$ we have
$\varphi_{t't''}(x_{t'})\le x_{t''}$.  When $\Lambda_t$ has enough points i.e.
if $I_t=\{t\}$, then the points in an open for the point topology would be
charactericed by $\{p,p\le [\chi_t]\}=U(\chi_t)$ for some $\chi_t\in X_t$.
When $\Lambda_t$ does not have enough points then  we have to modify the
definition of point spectrum and point topology correspondingly.  If
$x=(\ldots,x_t,\ldots)$ is $t$ accessible and $p_{t'}\in{\rm
Spec}(\Lambda_t,I_t)$ then we say $p_{t'}\in x$ if $t'\in J={\rm sup}(x)$ and
there exists an open $T$-interval $J_1\subset J$ with $t' \in J_1$ such that
for $t''\in J_1$ we have~: if $t'\le t''$ then
$p_{t''}=\varphi_{t't''}(p_{t'})\le x_{t''}$, or if $t''\le t'$ there is a
$p_{t''}\in\Lambda_{t''}$ such that
$\varphi_{t''t'}(p_{t''})=p_{t'},p_{t''}\le x_{t''}$, i.e. $\{p_{t''},t''\in
J_1\}$ is the restriction of a temporal point representing $p_{t'}$ defined
over a bigger $T$-interval $]t_0,t_1[$ containing both $t'$ and $t$ (note~:
$J_1$ need not contain $t$).
\subsection{Theorem}
The empty set together with the sets $U_t(x)=\{p_{t'},p_{t'}\in x$ for some
$t'\in I_t\}\subset {\rm Spec}(\Lambda_t,I_t)$, $x$ being $t$-accessible in
$\prod_{t\in T}\Lambda_t$, form a topology on ${\rm Spec}(\Lambda_t,I_t)$,
called {\bf spectral topology} at $t\in T$.
\paragraph{Proof}
Consider $x\neq y$ both $t$-accessible with respective $T$-intervals $J$,
resp. $J'$ contained in $I_t$.  If $p_{t'}\in U_t(x)\cap U_t(y)$ then $t'\in
J\cap J'$ and for every $t_1\in J$, $p_{t_1}\le x_{t_1}$, for every
$t_2\in J'$, $p_{t_2}\le y_{t_2}$.  Of course the interval $J\cap J'$ is
relative open around $t$.  If $t'\le t''$ with $t''\in J\cap J'$ then
$o_{t''}=\varphi_{t't''}(p_{t'})$ is idempotent in $\Lambda_{t''}$ because
$p_{t'}$ is in $\Lambda_{t'}$ as it is a point.  Hence we obtain~:
$$
p_{t''}=p_{t''}\wedge p_{t''}\le x_{t''}\wedge y_{t''}$$
Obviously for all $t''\le t'''$ in $J\wedge J'$ we do have~:
$\varphi_{t''t'''}(x_{t''}\wedge y_{t''})\le x_{t'''}\wedge y_{t'''}$.  On the
other hand, for $t''\le t'$ we obtain~: $\varphi_{t''t'}(p_{t''})=p_{t'}$ and
therefore $p_{t'}\le \varphi_{t''t'}(x_{t''})\le x_{t'}$, as well as $p_{t'}\le \varphi_{t''t'}(y_{t''})\le y_{t'}$.  Hence, again by idempotency of $p_{t'}$
in $\Lambda_{t'}$ we arrive at $p_{t'}\le x_{t'}\wedge y_{t'}$.  By restricting
$J\cap J'$ to the interval obtained by allowing only those $t''\le t'$ which
belong to an (open) unambiguity interval for $p_{t'}$ we arrive at a relative
open around $t$, say $J''\subset J\cap J'$, containing $t'$.
\par
Now for $p_{t''}$ with $t''\in J''$ it follows that $p_{t''}$ is idempotent
because both $p_{t''}$ and $p_{t''}\wedge p_{t''}$ map to $p_{t'}$ via
$\varphi_{t''t'}$ for $t''\le t'$ (other $t''$ in $J''$ are no problem).  Thus
for $t''$ in $J''$ we do arrive at $p_{t''}\le x_{t''}\wedge y_{t''}$.  Difine
$w$ by putting $w_{t''}=x_{t''}\wedge y_{t''}$ for $t''\in J''$.  Clearly, $w$
is $t$-accessible and $p_{t'}\in U_t(w)$.  Conversely if $p_{t'}\in U_t(w)$
then $p_{t'}\in U_t(x)\cap U_t(y)$ is clear because $J''$ used in the
definition of $w$ is open in $J\cap J'$.  Now we look at a union of
$U_{i,t}=U_t(x_i)$ for $i\in J$ and each $x_i$ being $A$-accessible with
corresponding relative open interval $J_i$ in $I_t$.  Define $w$ over the
``interval'' $J=\cup_i\{J_i,i\in {\cal J}\}$ by 
putting $w_t=\vee\{x_{i,t},i\in {\cal J}\}$ for $t\in J$.  It is clear that $J$
is relative open around $t$ and for all $t_1\le t_2$ in $J$ we have
$\varphi_{t_1t_2}(w_{t_1})\le w_{t_2}$ because $\varphi_{t,t_2}$ respects
arbitrary $\vee$.  Now $p_{t'}\in w$ means that $p_{t''}\le \vee
\{x_{i,t''},i\in {\cal J}\}$ for $t''$ in some relative open containiing $t'$,
say $J_1\subset J$.  We use relative open sets in $T$ because $I_t$ was closed
and there are two situations to consider concerning $t'\in I_t$.  First if
$t'$ is the lowest element of $I_t$ then for all $t''\in J_1$ we have that
$p_{t''}=\varphi_{t't''}(p_{t'})\le \varphi_{t't''}(\vee\{x_{i,t'},i\in {\cal
J}\})$ and for all $t'\le t_1\le t''$ we also obtain~: $p_{t_1}\le
\varphi_{t't_1}(\vee\{x_{i,t'},i\in {\cal J}\})$ and $p_{t''}\le
\varphi_{t_1t''}(\vee\{x_{t,t_1},i\in {\cal J}\})$.  Otherwise, if $t'$ is not
the lowest element of $I_t$ then we may restrict $J_1$ to be an open interval
$]t_0,t'_0[$ containing $t'$ with $t_0\in J$.  The same reasoning as in the
first case yields for all $t''\in ]t_0,t'_0[$ so that~: $p_{t''}\le 
\varphi_{t_0t''}(\vee\{x_{i,t_0},\tau\in{\cal J}\})$ and for any $t'\le t_1\le
t''$ $p_{t''}\le \varphi_{t_1t''}(\vee\{x_{i,t_1},i\in{\cal
J}\})=\vee\{\varphi_{t_1t''}(x_{i,t_1}),i\in{\cal J}\}$.  Since $t'\in J_1$ we
obtain $p_{t'}\le \vee\{\varphi_{t_1t'}(x_{i,t_1}),i\in J\}$ for all $t_1\in
[t_0,t']$.
\par
Since $p_{t'}$ is a point in $\Lambda_{t'}$ there is an $i_0\in J$ such that
$p_{t'}\le\varphi_{t_0t'}(x_{i_0,t_0})$ and therefore we have that $p_{t'}\le
\varphi_{t_1t'}(x_{i_0,t_1})$ with $t_1\in [t_0,t']$, the gain being that $i_0$
does not depend on $t_1$ here !  Now for $t''\ge t'$ in $J_1\cap J_{i_0}$
(note that this is not empty because $x_{i_0}$ is nonzero at $t_0$ because
$p_{t'}\le \varphi_{t_0t''}(x_{i_0,t_0})$ would then make $p_{t'}$ zero and we
do not look at the zero (the empty set) as a point of $\Lambda_t$) we obtain~:
$$
p_{t''}=\varphi_{t't''}(p_{t'})\le \varphi_{t't''}(x_{i_0,t'})\le
x_{i_0,t''}\leqno {(*)}$$
In the other situation $t''\le t'$ in $J_1\cap J_{i_0}$ we have
$\varphi_{t''t'}(p_{t''})=p_{t'},\varphi_{t''t'}(x_{i_0, t''})\le x_{i_0,t'}$.  
By restricting $J_1\cap J_{i_0}$ further so that the $t''\le t'$ are only
varying in an (open) unambiguity interval for $p_{t'}$, say $J_2\subset
J_1\cap J_{i_0}$, we arrive at one of two cases~: either $p_{t''}=x_{i_0,t''}$
or else $p_{t''}\neq x_{i_0,t''}$ and also $p_{t'}< x_{i_0,t'}$.  In the first
case $p_{t'}\in x_{i_0}$ follows because $p_{t_1}=\varphi_{t''t_1}(x_{i_0,
t''})\le x_{i_0,t_1}$ for $t_1$ in $]t'',1[\cap J_2$, the latter interval
containing $t'$ is relative open again.  In the second case we may look at
$p_{t'}< x_{i_0,t'}<1$, hence there exists a $z_{t''}$ such that
$p_{t''}<z_{t''}<1$ and $\varphi_{t''t'}(z_{t''})=x_{i_0,t'}$.  Again we have
to distinguish two cases, first $\varphi_{t''t'}(x_{i_0t''})=x_{i_0,t'}$ or
$\varphi_{t''t'}(x_{i_0,t''})<x_{i_0,t'}$.
\par
In the first case $z_{t''}$ and $x_{i_0,t''}$ map to the same element via
$\varphi_{t''t'}$, hence up to restricting the interval further such that
$t''$ stays within an unambiguity interval for $x_{i_0,t'}$, we may conclude
$z_{t''}=x_{i_0,t''}$ in this case and then $p_{t''}< x_{i_0,t''}$.  In the
second case we may look at~: $p_{t'}\le \varphi_{t''t'}(x_{i_0,t''})<
x_{i_0,t'}< 1$ (where the first inequality stems from (*) above).
Again restricting the interval further (but open) we find a $z'_{t''}$ in
$\Lambda_{t''}$ such that $x_{i,t''} < z'_{t''} < 1$ such that
$\varphi_{t''t'}(z'_{t''})=x_{i_0,t'}$.  Since we are dealing with the case
$p_{t''}\neq x_{i_0,t''}$ and we are in an unambiguity interval for $p_{t'}$
it follows that $p_{t'} <\varphi _{t''t'}(x_{i_0,t''})$.  Look at~:
$p_{t'}<\varphi_{t''t'}(x_{i_0,t''})< x_{i_0,t'}$ with
$\varphi_{t''t'}(p_{t''})=p_{t'}$ and $\varphi_{t''t'}(z'_{t''})=x_{i_0,t'}$;
by restructing the interval (open) further if necessary we obtain the
existence of $z''_{t''}$ such that, $p_{t''} < z''_{''}< z'_{t''}$ such that
$\varphi_{t''t'}(z''_{t''})=\varphi_{t''t'}(x_{i_0,t''})$.  Finally
restricting again the $t''\le t'$ to vary in an unambiguity interval for
$\varphi_{t''t'}(x_{i_0,t''})$ it follows that $z''_{t''}=xz_{i_0,t''}$ and
hence $z''_{t''}\ge p_{t''}$ yields $x_{i_0,t''}\ge p_{t''}$ for $t''$ in a
suitable relative open around $t$ containing $t'$.  This also in the case we
arrive at $p_{t'}\in x_{i_0}$ or $p_{t'}\in U_t(x_{i_0})$.  It follows that
${\cal U}_t(w)=U\{U_{i,t},i\in {\cal J}\}$ establishing that arbitrary
unions of opens are open.  By taking ${\cal U}_t(1)$ we obtain the whole
spectrum at $t$ as an open too.\hfill\qed
\subsection{A Possible Relation to $M$-Theory ?}
In noncommutative topology and derived point topologies the gen-topology
appears naturally (and it is a classical i.e. commutative topology (cf.
[VO1]).  Moreover continuity in the gen-topology also appears naturally in
noncommutative geometry of associative algebras but we did not ask the
$\varphi_{tt'}$ in the DNT-axioms to be continuous in the gen topology.
However one may prove that in general ``continuity of the $\varphi_{tt'}$ in
the gen-topologies of $\Lambda_t$ resp. $\Lambda_{t'}$ is an observed truth !
Consequently for $t'$ close enough to $t$ the $\varphi_{tt'}$ is continuous
with respect to the gen-topologies (cfr. [VO2]).
\par
In the mathematical theory all $\Lambda_t$ may be different and there is no
reason to aim at $Sl(\Lambda_t)$ nor ${\rm Spec}(\Lambda_t,I_t)$ to be
invariant under $t$-variation.  From the point of view of Physics one may
reason that only one case is important i.e. the case we see as `` reality''.
This being the utmost special case it is then not far-fetched to assume that
the dynamic noncommutative space has as commutative shadow the abstract
mathematical frame we reason in about reality. for example identified with 3
or 4 dimensional space or space-time, and also an observational mathematical
frame we measure in that is ${\rm Spec}(\Lambda_t,I_t)$ identified to an
$11$-dimensional space (for $M$-theory) or any other one fitting physical
interpretations in some theory one chooses to believe in.  An observed point
in ${\rm Spec}(\Lambda_t,I_t)$ is then  given by a string of elements say
$p_{t'}\in \Lambda_{t'}$, $t'\in J\subset I_t$ with $p_t\in\Lambda_t$ a
temporal point.  If $p_{t'}$ is a point then for all $t'\le t''$,
$\varphi_{t't''}(p_{t'})$ is idempotent so appears in the commutative shadow
$SL(\Lambda_{t''})$.  Hence an observed point in ${\rm Spec}(\Lambda_t,I_t)$
appears as a string in the base space $SL(\Lambda_{t''})(t'\le t'')$,
identified with $n$-dimensioal space but the string may ``start after'' $t$
when the point was ``observed''.  On the other hand, the assumption that the
system $(\Lambda_t,\varphi_{tt'},T\}$ is temporally pointed leads to a
decomposition of of every $p_{t'}$ into temporal points of $\Lambda_{t'}$
realizing it as an open of ${\rm Spec}[\Lambda_{t'},I_{t'}]$.  Thus in the
spectral space (identified with a certain $m$-dimensional space say), the
observed point appears as a ``string'' connecting opens i.e. a possibly more
dimensional string that can be thought of as a tube.  The difference between
the dimensions $m$ and $n$ has to be accounted for by the ``rank'' of $T$
(e.g. if $T$ were a group like $\mathbb{R}^d_+$, $d$ would be the rank) which
allowed to create the extra points in ${\rm Spec}(\Lambda_t,I_t)$ when
compared to $(SL(\Lambda_t)$.  Note that even when the $\varphi_{tt'}$ do not
necessarily define maps between ${\rm Spec}(\Lambda_t,I_t)$ and ${\rm
Spec}(\Lambda_{t'},I_{t'})$ or between $SL(\Lambda_t)$ and $SL(\Lambda_{t'})$, 
the given strings at the $\Lambda_t$-level do define sequences of elements or 
opens in the ${\rm Spec}(\Lambda_{t},I_t)$ resp. $SL(\Lambda_t)$ that may be 
viewed as strings, resp. tubes.  Two more intriguing observations~:
\begin{enumrom}
\item
Identifying ${\rm Spec}(\Lambda_t,I_t)$ to one fixed commutative world and
$SL(\Lambda_t)$ to another allows strings and tubes to be open or closed.
\item
Only temporal points corresponding to future points can be non-idempotent,
therefore all noncommutativity is due to future points and uncertainty may be
seen as an effect of the possibility that the interval needed  to realize the
temporal point $p_t$ by a point $p_{t''}\in\Lambda_{t''}$ for $t\le t''$ is
{\bf larger} than the unambiguity interval for $p_{t''}$ !  Passing from a
commutative frame $(SL(\Lambda))$ to noncommutative (dynamical) geometry and
phrasing theories and calculations in ${\rm Spec}(\Lambda_t,I_t)$ at the price
of having to work in higher dimension seems to fit quantum theories.  Of
course this is at the level of mathematical formalism, for suitable
interpretations within physics the physical entity connected to the notion of
point in ${\rm Spec}(\Lambda_t,I_t)$ 
should be the smallest possible, i.e. a kind of
building block of everything, so that observing it as a point in the moment
spaces is acceptable; that these points are mathematically described as
strings or higher dimensional strings via the noncommutative geometry is a
``Deus ex Machina'' pointing at an unsuspected possibility of embedding
$M$-theory in our approach.  No further speculation about this here, perhaps
specialists in string theory may be interested in investigating further this
formal incidence.
\end{enumrom}
\section{Moment Presheaves and Sheaves}
Continuing the point of view put forward in the short introduction to Section
2, points or more precisely functions defined in a set theoretic spirit,
should be replaced by a generalization of ``germs of functions'' obtained from
limit constructions in classical topology terms to noncommutative structures.
Thus the notion of point is replaced by an atavar of the notion ``stalk'' of a
pregiven sheaf, more correctly when different (pre)sheaves over a
noncommutative space are being considered, say with values in some nice
category of objects $\ul{\cal C}$, then a ``point'' is a suitable limit
functor on $\ul{\cal C}$-objects generalizing the classical construction of
localization (functor) at a point.  Assuming that a suitable topological space
and satisfactory sheaf of ``functions'' on it have been identified,
satisfactory in the sense that it allows to study the desired geometric
phenomena one is aiming at, then the notion of point via stalks should be
suitable too.  For example, prime ideals would be identified via stalks of the
structure sheaf of a commutative Noetherian ring without having to check a
primeness condition of the corresponding localization functor.  More on the
definition of noncommutative geometry via (localization) functors can be found 
in [VO1] where it is introduced as a functor geometry over a noncommutative
topologyee also [MVO] and [VOV].
\par
In this section we fix a category $\ul{\cal C}$ allowing limits and colimits;
we might restrict to Abelian or even Grothendieck categories but that is not
essential.  In fact, the reader who wants to fix ideas on a concrete situation
may choose to work in the category of abelian groups.
\par
For every $t\in T, \Gamma_t$ is a presheaf over $\Lambda_t$ and for $t\le t'$
in $T$ there is a $\phi_{tt'}:T_t\r T_{t'}$, defined by morphisms in $\ul{\cal
C}$ as follows~:
\begin{enumrom}
\item
For $\lambda_t\in\Lambda_t$ there is a
$\phi_{tt'}(\lambda_t):\Gamma_t(\lambda_t)\r
\Gamma_{t'}(\varphi_{tt'}(\lambda_t))$
\item
for $\mu_t\le \lambda_t$ in $\Lambda_t$ we have commutative diagrammes in
$\ul{\cal C}$~:
$$
\begin{diagram}
\Gamma_t(\lambda_t)\rto^{\phi_{tt'}(\lambda_t)}\quad
\dto^{\rho^{t}_{\lambda_t,\mu_t}}&
\qquad\Gamma_{t'}(\varphi_{tt'}(\lambda_t))
\dto^{\rho^{t'}_{\lambda_{t'},\mu_{t'}}}\\
\Gamma_t(\mu_t)\rto_{\phi_{tt'}(\mu_t)} \qquad &
\qquad\Gamma_{t'}(\varphi_{tt'}(\mu_t))\\
\end{diagram} 
$$
where we have written $\lambda_{t'},\mu_{t'}$ for $\varphi_{tt'}(\lambda_t)$,
resp. $\varphi_{tt'}(\mu_t)$ and $\rho^t_{\lambda_{t'}\mu_t}$ for the
restriction morphism of $\Gamma_t$.
\item
$\phi_{tt}(\lambda_t)=I_{\Gamma_t(\lambda_t)}$ for all $t\in T$, and for $t\le
t'$ and let $t'\le t''$ we have $\phi_{t't''}(\phi_{tt'}(T_t(\lambda_t)))=\phi_{tt''}(T_t(\lambda_t))$ for all $\lambda_t\in \Lambda_t$.
\end{enumrom}
The system $\{\Gamma_t,\phi_{tt'},T\}$ is called a (global) dynamical
presheaf over the DNT $\{\Lambda_t,\varphi_{tt'},T\}$.
\par Since sheaves on a noncommutative topology do not form a topos it is a
problem to define a suitable sheafification i.e. : can a presheaf $\Gamma$ on
$\Lambda$ be sheafified to a sheaf $\ul{\ul{a}}\Gamma$ on the same $\Lambda$
via a suitable notion of ``stalk'', then allowing interpretations in terms of
``points'' ?  In fact, the axioms of DNT allow to give a solution to the
sheafification problem at the price that the  sheaf $\ul{\ul{a}}\Gamma_t$ has
to be constructed over ${\rm Spec}(\Lambda_t,I_t)$~! 
\par From hereon we let $\ul{\Lambda}=\{\Lambda_t, \varphi_{tt'},T\}$ be a
temporally pointed system which is a space continuum.  We refer to $Y_t={\rm
Spec}(\Lambda_t,I_t)$ with its spectral topology as the {\bf moment space} at
$t\in T$.
\par
For $p_{t'}\in Y_t$ we may calculate (in $\ul{\cal C}$):
$\Gamma_{t',p_{t'}}=\mathop{\rm lim}\limits_{\longrightarrow}\Gamma_{t'}(x_t)$
where $\mathop{\rm lim}\limits_{\longrightarrow}$ is over $x_{t'}\in
\Lambda_{t'}$ such that $p_{t'}\le x_{t'}$, and where $x=(\ldots,x_t,\ldots)$
is $t$-accessible, in fact we have $p_{t'}\in x$.  In the foregoing we did not
demand $\ul{\Lambda}$ to derive from a system $\Xul$ and passing from $X_t$ to
$\Lambda_t$ as a generalized Stone space or pattern topology via $C(X_t)$,.
We preferred not to dwell upon the formal comparison
of dynamical theories for the $X_t$ and the $\Lambda_t$.  In order to keep
trace of a possible original $X_t$ if it was considered in the construction
of $\Lambda_t$ one may if desired use the following.
\subsection{Definition}
We say that $u_t\in\Lambda_t$ is {\bf classical} if $u_t=[\chi_t]$
for $\chi_t\in X_t$.  If $u_t$ is classical then there is an open interval
containing $t$ in $T$, say $L$, such that for every $t'\in L$ we have that
$u_{t'}$ is classical, where for $t\le t'$ we have $u_{t'}=\varphi_{tt'}(u_t)$
and for $t'\le t, u_{t'}$ is a suitably chosen representative for $u_t$,
$\varphi_{t't}(u_{t'})=u_t$.  Restricting further to an unambiguity interval of
$u_t$, the representations $u_{t'}$ for $t'$ in that interval are unique.
Since points of $X_t$ are by definition elements in $C(X_t)$, the
filter in $\Lambda_t$ defined by that point has a cofinal directed subset of
classical elements.  When in $\{\Lambda_t,\varphi_{tt'},T\}$ we restrict
attention to classical $x$, i.e. every $x_t$ classical $\Lambda_t$, then we
say that we look at a {\bf traditional system}.
\subsection{Lemma}
For a traditional space continuum with dynamical presheaf
$\{\Gamma_t,\phi_{tt'},T\}$, the stalk for $p_{t'}\in Y_t$ of $\Gamma_{t'}$ is
exactly $\Gamma_{t',p_{t'}}$ as defined above.
\paragraph{Proof}
In calculating $\mathop{\rm lim}\limits_{\longrightarrow}
\{\Gamma_{t'}(u_{t'}),p_{t'}\le u_{t'}\}$ we
may restrict to classical $u_{t'}$ in $\Lambda_{t'}$.  It now suffices to
establish the existence of a $t$-accessible $y$ such that $p_{t'}\in y$ and
$y_{t'}\le u_{t'}$.  From $p_{t'}\in U_t(x)$ we obtain $(\ldots,x_{t'},\ldots)$ with a relative open $T$-interval $J,t'\in J$, such that $p_{t''}\le x_{t''}$
for every $t''\in J$.  Since $u_{t'}$ and $x_{t'}$ are classical, so is
$u_{t'}\wedge x_{t'}$ and moreover $p_{t'}\le u_{t'}\wedge x_{t'}$ because
$p_{t'}$ is a point in $\Lambda_{t'}$ (hence idempotent !).  Let $J_1$ be an
open $T$-interval containing $t'$ such that $u_{t'}\wedge x_{t'}$ has a
representative $u_{t''}$ in $\Lambda_{t''}$ such that
$\varphi_{t''t'}=u_{t'}\wedge x_{t'}$.  Since $x_{t'}\neq u_{t'}\wedge x_{t'}$
may be assumed (otherwise put $y=x$) we arrive at $p_{t''}\le u_{t''}<
x_{t''}$.  Using the intersection of $J_1$ and the interval iaround $t'$
allowing to select classical $u_{t''}$, call this interval $J_2$, we put
$y_{t''}=u_{t''}$ for $t''\le t'$ in $J_2$ and $y_{t_1}=x_{t_1}$ for $t'<t_1$
in $J$.  Then $y$ is $t$-accesible with respect to the relative open
$T$-interval around $t$ just defined~: we have $y_{t'}\le u_{t'}$ and
$p_{t'}\in y$.  Consequently~:
$\mathop{\rm lim}\limits_{\mathop{\longrightarrow}\limits_{p_{t'} \le u_{t'}}}
\Gamma_{t'}(u_{t'}) = \mathop{\rm lim}\limits_{\mathop{\longrightarrow}
\limits_{p_{t'}\in x}}\Gamma_{t'}(x_{t'})$.
\hfill\qed\par
In the sequel we assume objects in $\ul{\cal C}$ are at least sets but let us
restrict to abelian groups.  Again let $\{\Gamma_t, \phi_{tt'},T\}$ be a
dynamical presheaf over a traditional space continuum.  On $Y_t$ we define a
presheaf, with respect to the spectral topology, by taking for 
${\cal P}(U_t(x))$ the abelian group in $\coprod_{t'\in I_t}\Gamma_{t'}(x_{t'})$formed by strings over ${\rm sup}(x)=\{t'\in I_t,x_{t'}\neq 0\}$. i.e. 
$\{\gamma_{t'}, t'\in {\rm sup}(x),\phi_{t''t'}(\gamma_{t''})=\gamma_{t'}$ for 
$t''\le t'$ in ${\rm sup}(x)\}$. 
Let us write $x<y$ if $x_{t'}\le y_{t'}$ for all $t'$ in $I_t$, in particular
$x < y$ means ${\rm sup}(x)\subset {\rm sup}(y)$.  In sheaf theory one usually
omits the empty set, so here the $o\in\Lambda_t$ at every $t$, and it makes
sense to do that here as well.  However one may define at every $t'\in T$,
$\Gamma_{t'}(0)=\mathop{\rm lim}\limits_{\longrightarrow}
\{\Gamma_{t'}(x_{t'}),x_{t'}$ classical in
$\Lambda_{t'}\}$ and all statements made in the sequel will remain
consistent.
\par
If $x<y$ then we have restriction morphisms $\rho^{t'}_{y_{t'},x_{t'}},
: \Gamma_{t'}(y_{t'})\r \Gamma_{t'}(x_{t'})$.
Commutativity of the diagrams in the beginning of the section yield
corresponding morphisms on the strings over the respective supports~:
$\rho_{y,x}:{\cal P}(U_t(y))\r {\cal P}(U_t(x))$.
For a point $p_{t'}$ we let $\eta(p_{t'})$ be the set of $U_t(x)$ such that we
have $p_{t'}\in U_t(x)$ i.e. $p_{t'}\in x$; in particular $t'\in J_x$ where
$J_x$ is the relative open around $t$ in the definition of $x$ and
consequently~: $t'\in \cap \{{\rm sup}(x),\eta(p_{t'})\ {\rm contains}\
U_t(x)\}$.  For the dynamical sheaf theory we may want to impose coherence
conditions on the system assuming some relations between $\Gamma_{t''}$ and
$\Gamma_{t'}$ if $t'$ and $t''$ are close enough in $T$.  We shall
restrict here to only one extra assumption, in some sense dual to the
unambiguity interval assumption for the underlying DNT.
\subsection{Definition}
The dynamical presheaf $\{\Gamma_t,\phi_{tt'},T\}$ on a traditional space
continuum is locally temporaly flabby at $t\in T$ if for $t$-accessible $x$
such that $p_{t'}\in x$ and $s_{t'}\in\Gamma_{t'}(x_{t'})$ there exists a
$t$-accessible $y<x$ with $p_{t'}\in y$ and a string $\sol\in {\cal P}(U_t(y))$
such that $\sol_{t'}=\rho^{t'}_{x_{t'},y_{t'}}(s_{t'})$.
\subsection{Theorem}
To a dynamical presheaf on a traditional space continuum there corresponds for
every $t\in T$ a presheaf ${\cal P}_t$ on the moment space ${\rm
Spec}(\Lambda_t,I_t)$ with its spectral topology given by the $U_t(x)$ for
$t$ accessible $x$.  In case all $\Gamma_{t'},t'\in I_t$, are separated
presheaves then ${\cal P}_t$ is separated too.  The sheafification 
$\ul{\ul{a}}{\cal P}_t$ of ${\cal P}_t$ on the moment space 
${\rm Spec}(\Lambda_t,I_t)=Y_t$ is called the {\bf moment sheaf} of 
{\bf spectral sheaf} at $t\in T$.  In case the dynamical
presheaf is locally temporally flabby (LTF) then for any point $p_{t'}\in Y_t$
the stalk ${\cal P}_{t,p_{t'}}$ may be identified with $\Gamma_{t',p_{t'}}$.
\paragraph{Proof}
At every $t\in T$, ${\cal P}_t$ is the spectral presheaf constructed on ${\rm
Spec}(\Lambda_t,I_t)$ with its spectral topology.  Now suppose all $\Gamma_t$
are separated presheaves and look at a finite cover
$U_t(x)=U_t(x_1)\cup\ldots\cup U_t(x_n)$ and a $\gamma\in\Gamma_t(U_t(x))$
such that for $i=1,\ldots,n$, $\rho_{x.x_i}(\gamma)=0$.  We have seen before
that the union $U_t(x_1)\cup\ldots\cup U_t(x_n)$ corresponds to the $t$
accessible element $x_1\wedge\ldots\wedge x_n$ obtained as the string over ${\rm
sup}(x_1)\wedge\ldots\wedge{\rm sup}(x_n)$ given by the $x_{1,t'}\cup\ldots\cup
x_{n.t'}$ in $\Lambda_{t'}$.  For all $t'\in {\rm sup}(x)$ we obtain, in view
of the compatibility diagrams for restrictions and $\phi_{t't''},t'\le
t^n:\rho^{t'}_{x_{t'},x_{i,t'}}(\gamma_{t'})=0$, for $i=1,\ldots,n$. The assumed
separatedness of $\Gamma_{t'}$, for all $t'$ then leads to $\gamma_{t'}=0$ for
all $t'\in{\rm sup}(x)$ and therefore $\gamma=0$ as a string over ${\rm
sup}(x)$.  Consequently ${\cal P}_t$ is separated, for all $t\in T$.  In order 
to calculate the stalk at $p_{t'}\in{\rm Spec}(\Lambda_t,I_t)$ for ${\cal P}_t$ 
we have to calculate~: $\mathop{\rm lim}\limits_{\mathop{\longrightarrow}
\limits_{p_{t'}\in x}}\Gamma_t(U_t(x))=E_{t'}$.
\par
Starting with $p_{t'}\in x$ for some $t$-accessible $x$ we have a
representative $\gamma_x\in \Gamma_t(U_t(x))$ being a string over ${\rm
sup}(x)$ and the latter containing a relative open $J(x)$ around $t$
containing $t'$.  So an element $e_{t'}$ in $E_{t'}$ may be viewed as given by
a direct family $\{\gamma_x,p_{t'}\in x, \rho_{x,y}(\gamma_x)=\gamma_y\
{\rm for}\ y<x\}$.  At $t'$, which is in ${\rm sup}(x)$ for all $x$ appearing
in the forgoing family (as $U_t(x)$ varies over $\eta(p_{t'}))$,
we obtain $\{(\gamma_x)_{t'}, p_{t'}\le
x_{t'},\rho^{t'}_{x_{t'},y_{t'}}((\gamma_x)_{t'})=(\gamma_y)_{t'}$ which
defines an element of $\Gamma_{t',p_{t'}}$, say $\eol_{t'}$.  We have a
well-defined map $\pi_{t'}:E_{t'}\r \Gamma_{t',p_{t'}},e_{t'}\mapsto
\eol_{t'}$.  Without further assumptions we therefore arrive at a sheaf
$\ul{\ul{a}}{\cal P}_t$ with stalk $E_{t'}$ at $p_{t'}$ and a presheaf map 
${\cal P}_t\r\ul{\ul{a}}{\cal P}_t$ which is ``injective'' in case all 
$\Gamma_{t''}$ are separated.  Now we have to make use of the locally 
temporally flabbyness (LTF).  Look at a germ $s_{t'}\in (\Gamma_{t'})_{p_{t'}}$.
In view of Lemma 3.2. there exists a $t$-accessible $x$ such that
$s_{t'}\in\Gamma_{t'}((x_{t'})$ with $p_{t'}\in x$, in particular $p_{t'}\le
x_{t'}$.
\par
The LTF-condition allows to select a $t$-accessible $y<x$ with $p_{t'}\in y$
together with a string, $\vec{s(y)}\in {\cal P}(U_t(y))$ such that 
$\vec{s_{t'}}(y)=\rho^{t'}_{x_{t'},y_{t'}}(s_{t'})$.  The element $e_{t'}$
in $E_{t'}$ defined by the directed family obtained by taking restrictions of
$\vec{s_{t'}}(y)$ has $\eol_{t'}$ exactly $s_{t'}$ (note that $t'$
supports all the restrictions of $\vec{s_{t'}}(y)$ because $y$ varies in
$\eta(p_{t'})$).  Thus $\pi_{t'}:E_{t'}\r \Gamma_{t',p_{t'}}$ is epimorphic.
If $e_{t'}$ and $e'_{t'}$ have the same image under $\pi_{t'}$ then there is a
$t$-accessible $y$ such that $e_{t'}-e'_{t'}$ is represented by the
zero-string over ${\rm sup}(y)$; in fact this follows by taking $s_{t'}=0$ in
the foregoing.  Leading to a $t$-accessible $y$ as above that may be
restricted to a $t$-accessible $y'$ defined by taking for ${\rm sup}(y')$ the
relative open $J$ containing $t'$ in the support of $y$ where 
$\vec{s_{t'}}(y)=0$.  Therefore, $\pi_{t'}$ is also injective.\hfill\qed
\par
Can one avoid a condition like $LTF$ in the foregoing theorem ?  It seems that
the idea of ``germ'' appearing in the notion of stalks spatially needs an 
extension in the temporal direction, so probably some condition close to $LTF$
is really necessary here.
\section{Some Remarks on Spectral families and Observables}
Let $\Gamma$ be any totally ordered abelian group.  On a noncommutative
topology $\Lambda$ we define a {\bf $\Gamma$-filtration} by a family
$\{\lambda_{\alpha},\alpha\in\Gamma\}$ such that for $\alpha\le \beta$ in
$\Gamma$, $\lambda_{\alpha}\le \lambda_{\beta}$ in $\Lambda$ and
$\vee\{\lambda_{\alpha},\alpha\in\Gamma\}=1$.  A $\Gamma$-filtration is said
to be separated if from $\gamma={\rm inf}\{\gamma_{\alpha},\alpha\in{\cal
A}\}$ in $\Gamma$ it follows that
$\lambda_{\gamma}=\wedge\{\lambda_{\alpha},\alpha\in{\cal A}\}$ and
$0=\wedge\{\lambda_{\gamma},\gamma\in\Gamma\}$.  {\bf A $\Gamma$-spectral
family in $\Lambda$} is just a separated $\Gamma$-filtration, it may be seen
as a map $F:\Gamma\r \Lambda,\gamma\mapsto \lambda_{\gamma}$, where $F$ is a
poset map satisfying the separatedness condition.  Note that by definition the
order in $\wedge \{\lambda_{\gamma},\gamma\in \Gamma\}$ does not matter but
$\lambda_{\gamma_{\alpha}}$ need not be idempotent in $\Lambda$.  Taking
$\Gamma=\mathbb{R}_+$ and $\Lambda=L(H)$ the lattice of a Hilbert space $H$,
we recover the usual notion of a spectral family.  We say that a
$\Gamma$-spectral family on $\Lambda$ is idempotent if
$\lambda_{\gamma}\in{\rm id}_{\wedge}(\Lambda)$ for every $\gamma\in\Gamma$.
\paragraph{Observation}
If $\Gamma$ is indiscrete, i.e. for all $\gamma\in\Gamma$,
$\gamma={\rm inf}\{\tau,\gamma < \tau\}$ (example $\Gamma=\mathbb{R}^n_+$),
then every $\Gamma$-spectral family is idempotent.
\subsection{Proposition}
\begin{enumrom}
\item
Let us consider a $\Gamma$-spectral family on $\Lambda$, then for
$\gamma,\tau\in\Gamma:\lambda_{\gamma}\wedge
\lambda_{\tau}=\lambda_{\tau}\wedge\lambda_{\gamma}=\lambda_{\delta}$, where
$\delta={\rm min}\{\tau,\gamma\}$.
\item
If the $\Gamma$-spectral family is idempotent then for $\gamma,\tau\in
\Gamma$,
$\lambda_{\gamma}\wedge\lambda_{\tau}=\lambda_{\tau}\wedge\lambda_{\gamma}$ and
the $\Gamma$-spectral family on $\Gamma$ is in fact a $\Gamma$-spectral family
of the commutative shadow $SL(\Lambda)$.
\end{enumrom}
\paragraph{Proof}
Easy enough.\hfill\qed
\par
A filtration $F$ on a noncommutative $\Lambda$ is said to be {\bf right
bounded} if $\lambda_{\gamma}=1$ for some $\gamma\in\Gamma$, $F$ is {\bf left
bounded} if $\lambda_{\delta}=0$ for some $\delta\in\Gamma$.
\par
For a right bounded $\Gamma$-filtration $F:\Gamma\r \Lambda$ we may define for
every $\mu\in\Lambda$ the induced filtration $F|_{\mu}:\Gamma\r \Lambda(\mu)$
where we use $\mu=1_{\wedge(\mu)}$,
$\Lambda(\mu)=\{\lambda\in\Lambda,\lambda\le\mu\}$.  Note that $F|_{\mu}$ need
{\bf not} be separated whenever $F$ is, indeed if $\delta={\rm
inf}\{\delta_{\alpha},\alpha\in A\}$ in $\Gamma$ then
$\lambda_{\delta}=\wedge\{\lambda_{\delta_{\alpha}},\alpha\in {\cal A}\}$ in
$\Gamma$ then
$\lambda_{\delta}=\wedge\{\lambda_{\delta_{\alpha}},\alpha\in{\cal A}\}$ but
$\mu \wedge\lambda_{\delta}$, and
$\wedge\{\mu\wedge\lambda_{\alpha},\alpha\in{\cal A}\}$ need not be lequal in
general.
\subsection{Proposition}
If $F$ defines a right bounded $\Gamma$-spectral family on $\Lambda$ then
$F|_{\mu}$ is a spectral family of $\Lambda(\mu)$ in each of the following
cases~:
\begin{itemize}
\item[a.]
$\mu\in{\rm id}_{\wedge}(\Lambda)$ and $\mu$ commutes with all
$\lambda_{\alpha},\alpha\in\Gamma$.
\item[b.]
$\mu\wedge\lambda_{\alpha}$ is idempotent for each $\alpha\in\Gamma$.
\end{itemize}
\paragraph{Proof}
Easy and straightforward.\hfill\qed
\smallskip\par
An element $\mu$ with property a. as above is called an {\bf $F$-centralizer}
of $\Lambda$.
\subsection{Corollary} In case $\Lambda$ is a lattice then for every
$\mu\in\Lambda$ a right bounded $\Gamma$-spectral family of $\Lambda$ induces
a right bounded spectral family on $\Lambda(\mu)$.\par
Let $F$ be a $\Gamma$-spectral family on a noncommutative $\Lambda$.  To
$\lambda\in\Lambda$ associate $\sigma(\lambda)\in \Gamma\cup \{\infty\}$ where
$\sigma(\lambda)={\rm inf}\{\gamma,\lambda\le\lambda_{\gamma}\}$ and we agree
to write ${\rm inf}\phi=\infty$.  The map $\sigma:\Lambda\r
\Gamma\cup\{\infty\}$ is a generalization of the principal symbol map in the
theory of filtered rings and their associated graded rings.  We refer to
$\sigma$ as the {\bf observable function} of $F$.
$$
\begin{array}{r}
{\rm Clearly~} : \sigma(\lambda\wedge \mu)\le {\rm
min}\{\sigma(\lambda),\sigma(\mu)\}\\
\sigma(\lambda\vee\mu)\le {\rm max}\{\sigma(\lambda),\sigma(\mu)\}\\
\end{array}
$$
The {\bf domain of $\sigma$} is $\cup\{[0,\lambda_{\gamma}],\gamma \in \Gamma\}$
and observe that $\vee\{\lambda_{\gamma},\gamma\in\Gamma\}=1$ does not imply
that $D(\sigma)=\Lambda$ (can even be checked for
$\Gamma=\mathbb{R}_+,\Lambda=L(H)$). 
\par
If $F:T\r L(H)$ is a $\Gamma$-spectral family and $V\subset H$ a linear
subspace, then we may define $\gamma_{\vee}\in\Gamma$, $\gamma_{V}={\rm
inf}\{\gamma\in\Gamma,\subset L(H)_{\gamma}\}$, again putting ${\rm
inf}\phi=\infty$.  The map $\rho:L(H)\r \Gamma\cup\{\infty\}, U\mapsto
\rho(U)=\gamma_U$ is well-defined.  One easily verifies for $\cup$ and $\vee$
in $H$~:
$$
\begin{array}{l}
\rho(U+V)\le {\rm max}\{\rho(U),\rho(V)\}\\
\rho(U\cap V)\le {\rm min}\{\rho(U),\rho(V)\}\\
\end{array}
$$
\sp
The function $\rho$ defines a $\ul{\rho}$ defined on $H$ by putting
$\rho(x)=\rho(\mathbb{C}x)$.  We denote $\ul{\rho}$ again by $\rho$ and call
it the {\bf pseudo-place of the $\Gamma$-spectral family}.  Then any
$\Gamma$-spectral family defines a function on the projective Hilbert space 
$\mathbb{P}(H)$ described on the lines in $H$ by $\ol{\rho}:\mathbb{P}(H)\r
\Gamma,\ul{\mathbb{C}v}\mapsto \rho(\mathbb{C}v)$, where we wrote
$\ul{\mathbb{C}v}$ for $\mathbb{C}v$ as an object in $\mathbb{P}(H)$.
\par
The pseudo-place aspect of $\rho$ translates to $\ol{\rho}$ in the following
sense~: $\mathbb{C}w\subset \mathbb{C}v+\mathbb{C}u$ we have
$\ul{\rho}(\ul{\mathbb{C}w})\le {\rm max}\{\ol{\rho}(\ul{\mathbb{C}v}),\ol{\rho}(\ul{\mathbb{C}u})\}$.
\par
A linear subspae $U\subset H$ such that $\mathbb{P}(U)\subset
\ol{\rho}^{-1}(]-\infty,\gamma])$ allows for $u\neq 0$ in
$U:\rho(\mathbb{C}u)\le \gamma$ i.e. $u\in L(H)_{\gamma}$. Hence, the largest
$U$ in $H$ such that $\mathbb{P}(U)$ is in $\ol{\rho}^{-1}(]-\infty,\gamma])$
is exactly $L(H)_{\gamma}$; this means that the filtration $F$ may be
reconstructed from the knowledge of $\ol{\rho}$.  One easily recovers the
classical result that maximal abelian Von Neumann regular subalgebras of
${\cal L}(H)$ correspond bijectively to maximal distributive lattices in
$L(H)$.  Since any $\Gamma$-spectral family is a directed set in $\Lambda$ it
defines an element of $C(\Lambda)$ which we call a {\bf $\Gamma$-point}.  The
set of $\Gamma$-points of $\Lambda$ is denoted $[\Gamma]\subset C(\Lambda)$.
We may for example think of $[\mathbb{R}]\subset C(L(H))$ as being identified
via the Riemann-Stieltjens integral to the set of self-adjoint operators on
$H$.
\par
Let $\sigma:\Lambda\r \Gamma\cup\{\infty\}$ be the observable functions of a
$\Gamma$-spectral family on $\Lambda$ defined by $F:\Gamma\r\Lambda$.  Put 
${\cal F}:C(\Lambda)\r \Gamma\cup\{\infty\}$,$[A]\mapsto{\rm
inf}\{\gamma\in\Gamma,\lambda_{\gamma}\in\Aol\}$, $\Aol$ the filter of $A$.
Then $\widehat{\sigma}$ is the observable corresponding to the
$\Gamma$-filtration on $C(\Lambda)$ defined by $[A]_{\gamma}$, where for
$\gamma\in\Gamma$,$[A]_{\gamma}$ is the class of the smallest filter containing
$\lambda_{\gamma}$ i.e. the filter $\{\mu\in\Lambda,\lambda_{\gamma}\le\mu\}$.
This is clearly a $\Gamma$-spectral family because in fact
$[A]_{\gamma}<[\lambda_{\gamma}]$.  We define $[\Gamma]\cap {\rm Sp}(\Lambda)
=\Gamma{\rm -Sp}(\Lambda),[\Gamma]\cap Q Sp(\Lambda)[\Gamma]\cap QSp(\Lambda)
=\Gamma-QSp(\Lambda)$ and similarly with $p$
replaced by $P$ when ${\rm Id}_{\wedge}(C[\Lambda])$ is considered instead of
${\rm id}_{\wedge}(C(\Lambda_-))$ (see section 1).
\par
In view of Proposition 4.1.i. a $\Gamma$-spectral family is contained in a
sublattice (that is with commutative $\wedge$) of the noncommutative
$\Lambda$, in fact $\{\lambda_{\gamma},\gamma\in\Gamma\}$ is such a
sublattice.  If $Ab(\Lambda)$ is the set of maximal commutative sublattices of
$\Lambda$ then every $\Gamma$-spectral family in $\Lambda$ is a
$\Gamma$-spectral family in some $B\in Ab(\Lambda)$ ($B$ refers to Boulean
sector in case $\Gamma=\mathbb{R}_+,\Lambda=L(H)$).  The above remarks may be
seen as a generalization of the result concerning maximal commutative Von
Neumann regular subalgebras in ${\cal L}(H)$ quoted above.
\par
$\Gamma$-spectral families may be defined on the moment spaces ${\rm
Spec}(\Lambda_t,T_t)$ in exactly the way described above as filtrations
$\{U_t(x_{\gamma}),\gamma\in\Gamma\}$, where each $x_{\gamma}$
is $t$-accessible, defining a separated $\Gamma$-filtration.  For $t''\in I_t$
we may look at $V_t(x_{\gamma})=\{p_{t''},p_{t''}\in U_t(x_{\gamma})\}$, agaiin
$p_{t''}=\varphi_{t't''}(p_{t'})$ or $\varphi_{t't''}(p_{t'})=p_{t''}$
depending whether $t'\le t''$ or $t''\le t'$.  The family
$\{V_t(x_{\gamma}),\gamma\in\Gamma\}$ need not (!) be a $\Gamma$-spectral
family at $t''\in T$.  A stronger notion of {\bf dynamical spectral family}
may be obtained by demanding the existence of stringwise spectral families in a
relative open $T$-interval $J$ around $t$.  Then indeed at $t''\in J\subset
I(t)$ such a stringwise $\Gamma$-spectral family induces a $\gamma$-spectral
famly in $\Lambda_{t''}$ but not immediately on ${\rm
Spec}(\Lambda_{t''},I_{t^n})$ unless a more stringent relation is put on
$I_{t^n}$ and its comparison with respect to $I_t$.  We just point out the
interesting problems arising with respect to observables when passing to
moment spaces but this is work in progress.
\section*{References}
\begin{itemize}
\item[[VO1]]
F. Van Oystaeyen, {\sl Algebraic Geometry for Associative Algebras}, M.
Dekker, Math. Monographs, Vol. 232, New York, 2000.
\item[[Co]]
A. Connes, {\sl $C^*$-Alge\`ebres et G\'eom\'etries Differentielle}, C. R.
Acad. Sc. Paris, 290, 1980, 599-604.
\item[[CoD]
A. Connes, M. Dubois-Violette, {\sl Noncommutative Finite Dimensional
Manifolds I, Spherical Maniforls and related Examples}, Notes IHES, MO1, 32,
2001.
\item[[A.T.V.]]
M. Artin, J. Tate, M. Van den Bergh, {\sl Modules over Regular Algebras of
Dimension 3}, Invent. Nath. 106, 1991, 335-388.
\item[[VdB]]
M. Van den Bergh, {\sl Blowing up of Noncommutative Smooth Surfaces}, Mem.
Amer. Math. Soc., 154, 2001, no. 734.
\item[[Ko]]
M. Kontsevich, {\sl Deformation, Quantization of Poissin Manifolds}, q-alg,
9709040, 1997.
\item[[VO.2]]
F. Van Oystaeyen, {\sl Virtual Topology amd Functor Geometry}, lecture Notes
UA, Submitted.
\item[[MVO]]
D. Murdoch, F. Van Oystaeyen, {\sl Noncommutative Localization and Sheaves,}.
J. of Algebra, 35, 1975, 500-525.
\item[[S]]
J.-P. Serre, {\sl Faisceaux Alg\'ebriques Coh\'erents}, Ann. Math. 61, 1955,
197-278.
\item[[VOV]]
F. Van Oystaeyen, A. Verschoren, {\sl Noncommutative Algebraoc Geometry}, LNM
887, Springer Verlag, berlin, 1981.
\end{itemize}
\end{document}